\documentclass[12pt]{article}

\usepackage{hyperref}

\begin{document}

\hfill 

\hfill 

\bigskip\bigskip

\begin{center}

{{\Large\bf  Long distance regularization in chiral perturbation theory with
  decuplet fields}}

\end{center}

\vspace{.4in}

\begin{center}
{\large B.~Borasoy$^a\footnote{email: borasoy@physik.tu-muenchen.de}$,
 B.~R.~Holstein$^b\footnote{email: holstein@physics.umass.edu}$,
 R.~Lewis$^c\footnote{email: randy.lewis@uregina.ca}$
 and P.-P.~A.~Ouimet$^c\footnote{email: ouimet1p@uregina.ca}$}

\bigskip

\bigskip

a) \href{http://www.ph.tum.de/}{Physik Department},
Technische Universit{\"a}t M{\"u}nchen,\\
D-85747 Garching, Germany \\
b) \href{http://www.physics.umass.edu/}{Department of Physics},
University of Massachusetts,\\
Amherst, MA 01003, USA \\
c) \href{http://www.phys.uregina.ca/}{Department of Physics},
University of Regina, \\
Regina, SK, S4S 0A2, Canada

\vspace{.2in}

\end{center}

\vspace{.5in}

\thispagestyle{empty} 

\begin{abstract}
We investigate the use of long distance regularization in $SU(3)$
baryon chiral perturbation theory with decuplet fields. The one-loop
decuplet contributions to the octet baryon masses, axial couplings, S-wave nonleptonic
hyperon decays and magnetic moments are evaluated in a chirally consistent
fashion by employing a cutoff to implement long distance regularization. 
The convergence of the chiral expansions of these quantities is improved 
compared to the dimensionally regularized version which indicates
that the propagation of Goldstone bosons over distances smaller
than a typical hadronic size, which is beyond the regime of chiral perturbation
theory but included by dimensional regularization, is removed by use of a
cutoff.
\end{abstract}\bigskip

\vfill

\section{Introduction}
The holy grail of particle/nuclear physics is to be able to make direct
contact between experimental reality and the QCD Lagrangian,
\begin{equation}
{\cal L}_{QCD}=-{1\over 2}{\rm Tr}G_{\mu\nu}G^{\mu\nu}+\bar{q}
(i\not\!\!{D}-m)q\label{eq:qcd},
\end{equation}
with
\begin{equation}
iD_\mu=i\partial_\mu-gA_\mu,\qquad G_{\mu\nu}=\partial_\mu A_\nu
-\partial_\nu A_\mu-gA_\mu\times A_\nu,
\end{equation}
which is presumed to underlie them.  At the very highest energies the 
validity of asymptotic freedom permits this confrontation via perturbative
QCD, since hard scattering is able to access the basic quark/gluon structure 
of hadronic systems.  However, at low energies the quark/gluon degrees of 
freedom in terms of which ${\cal L}_{QCD}$ is written are hidden within
hadronic structure and any solution is inherently nonperturbative.  In order 
to make progress we must therefore
\begin{itemize}
\item[i)] resort to models;
\item [ii)] attempt an ``exact'' solution via lattice gauge techniques;
\item [iii)] exploit the (broken) chiral symmetry of Eq.~(\ref{eq:qcd})
via chiral perturbation theory.
\end{itemize}
The use of models has obvious limitations in that, despite being able to
represent data, the size and scope of possible corrections is difficult 
to assess.  Lattice gauge theory, while promising, is presently limited at
least as far as reasonably precise calculations are concerned to two and three
point functions, while experiments obviously do not have such limitations.
The use of chiral perturbative methods\cite{gl} on the other hand has been enormously
successful over a broad range of applications in the case of Goldstone
bosons and their weak and electromagnetic interactions at low 
energy\cite{hol}.  In
the case of baryons, things are not so simple.  For one thing, since the
mass of the nucleon --- $M$ --- is comparable to the chiral 
scale --- $\Lambda_\chi\sim
4\pi F_\pi$ --- retention of strict power counting requires an additional
expansion in terms of $1/M$ which must be accomplished by a Foldy-Wouthuysen
transformation or equivalent technique.  This method is called heavy 
baryon chiral perturbation
theory (HB$\chi$pt)\cite{hb} and has been applied to a large variety of baryonic
and baryon-meson interactions\cite{me}.  However, things are not as simple as in the
meson case.  On the one hand, there
exist reactions such as low energy pion photoproduction and Compton 
scattering wherein there has been a generally very successful interplay of 
near threshold experiments and HB$\chi$pt predictions.  However, there 
have also been failures and this is due to convergence problems in  
the chiral expansion.  This slow convergence is due for one thing 
to the feature that in the baryon sector the expansion is linear in 
$(q,m_\pi)/\Lambda_\chi$ rather than quadratic as in the case of the mesons,
but it also seems to have a deeper origin.  In any case, even more serious 
problems arise when one attempts to extend
this formalism to chiral $SU(3)$ by inclusion of kaon and eta degrees
of freedom.
In this case, since the leading chiral corrections
generally go as the meson mass to a positive power (with or without an
accompanying logarithm) any corrections from kaons or etas are quite
substantial and often end up destroying the generally good agreement between 
experiment and the venerable and successful predictions of $SU(3)$.
Examples of this phenomenon include
\begin{itemize}
\item [i)] baryon masses, wherein the success of the simple Gell-Mann-Okubo
mass formula at first order in $SU(3)$ breaking\cite{gmo} 
picks up very large corrections
proportional to $m_P^3$ at one loop in HB$\chi$pt\cite{bkm};
\item [ii)] axial couplings in semileptonic hyperon decays, 
wherein the success of the lowest order $SU(3)$
representation in terms of F,D couplings\cite{fad} is modified subtantially by
${\cal O}(m_P^2\log m_P^2)$ effects at one loop\cite{ola};
\item [iii)] S-wave amplitudes in nonleptonic hyperon decays, wherein
the remarkable success of the lowest order current algebra/PCAC results
which utilize a simple f,d parameterization\cite{cap} 
are modified at one loop by
${\cal O}(m_P^2\log m_P^2)$ effects\cite{ola};
\item [iv)] baryon magnetic moments, wherein the basic success of a simple
$SU(3)$ parameterization and its simple quark model extension\cite{bmm} 
is modified by
linear terms in $m_P$ which arise at one loop\cite{olm}.
\end{itemize}     
Of course, these are not fatal problems and can be cured by counterterm
corrections at higher order in the chiral expansion.  The exact way in which
this occurs has been seen in ${\cal O}(p^4)$ calculations
which have been done in the case of the mass terms and of the magnetic 
moments\cite{mbm, mml}.  
However, to the extent that such corrections are large
they indicate that the series which describes the chiral corrections 
is of the generic form
\begin{equation}
{\rm Amp}=A_0(1-1+1-1+\ldots)
\end{equation}
which is {\it not} what one looks for in an {\it effective} chiral expansion.

In previous papers we have suggested a solution to this dilemma --- a way
to make the effective field theory more effective\cite{ldr}.  The 
issue in any effective field theory treatment is to make sure that all the
relevant scales are accounted for and we have argued that in the conventional
HB$\chi$pt analysis this is {\it not} the case, in that the size of the
baryon is not included.  This can be seen for one thing from the feature 
that in the usual dimensional regularization (dim reg)
prescription, by which chiral loops are regularized,
contributions from high and low energy scales are weighted in the same
fashion.  (It should be noted that this same equal weighting has been shown to
lead to other difficulties in understanding the OPE scaling of 
operators --- {\it e.g.} operators which are purportedly of higher
dimension contribute at lower order than expected on dimensional
grounds\cite{don}.)
On the other hand {\it physics} considerations tell us that, while
components of the chiral Lagrangian which are large compared to the
baryon size can be trusted to be determined by the underlying chiral symmetry
of the problem, this is certainly not true for pieces which are small 
compared to the baryon radius and which are certainly modified by 
hadronic structure.  Our solution is to include some sort of form factor
in such calculations, which is unity at large distances ({\it i.e.} $r>>R_B$)
but which vanishes for distances short compared to the baryon radius.  
For calculational simplicity we have employed a
dipole form factor
\begin{equation}
F(q^2)=\left({\Lambda^2\over \Lambda^2-q^2}\right)^2
\end{equation}
with a cutoff parameter in the range 300 MeV $<\Lambda<$ 600 MeV, but the
precise shape of such a function is irrelevant --- any function which
has the right physics should suffice.  For example, we have
shown that the same physics results from use of lattice size as a 
regulator\cite{rpo}.  
The purpose of the cutoff function is to remove the model-dependent short
distance portions of the loop integrals which are not suppressed in 
dim reg.
For obvious reasons we have 
called this procedure ``long distance regularization'' (LDR)
 and have demonstrated that
it is guaranteed to isolate the basic chiral physics of any baryonic
process in as model-independent a fashion as possible without the
unphysical equal weighting of long and short distance physics.  In 
particular we have shown explicitly how the power law dependences
on $\Lambda$ which arise in this procedure can be absorbed into
phenomenologically-determined
counterterms, while the remaining chiral corrections do not
appreciably modify the successful lowest order $SU(3)$ predictions provided 
that the cutoff is chosen in the range given above.  Also, we have
found that there exists only weak dependence on the cutoff $\Lambda$.
If one worked to all chiral orders,
all dependence could be absorbed by the LECs and the theory would not
depend on the value of Lambda. Since we work at finite order, the residual
dependence is expected to occur beyond the order we are working, so that the
dependence on $\Lambda$ should be rather mild.

That this idea makes sense physically is also indicated by the fact that it
matches nicely onto successful and chirally well motivated model-dependent
approaches such as the cloudy bag model\cite{clb} or a recent Bethe-Salpeter
approach to baryon structure\cite{kre}.  For example, in the dynamical model
advocated in the latter picture, the chiral self-energy
corrections to the baryon masses are of the form
\begin{equation}
\Sigma_N(M_N)\propto \int_0^\infty dk\left[{k^4u^2(k)\over \omega^2(k)}
+{32\over 25}{k^4u^2(k)\over \omega(k)(M_\Delta-M_N-\omega(k)}\right],
\end{equation}   
where $\omega(k)=\sqrt{\vec{k}^2+m_\pi^2}$, 
which has an identical form to that found in the LDR
picture provided that we make the identification
$$F(k^2)=u^2(k).$$
Now in the dynamical model employed in Ref.~\cite{kre} the form of the
function $u(k)$ is generated dynamically and there is no freedom
to choose the shape.  Nevertheless, what is found matches nicely onto 
our assumed dipole form provided one
picks a cutoff in the vicinity $\Lambda\sim 500$ MeV.  Likewise similar
effects are found in any sort of cloudy bag model approach, as argued by
Leinweber et.~al.\cite{lei}.

A further reason for and application of such calculations was advocated by
the Adelaide group\cite{lei},\cite{you},\cite{wri}, 
who emphasize that such realistic chiral baryon 
corrections are just what is needed in order to extrapolate state of the
art lattice calculations, which are done for quark masses (and therefore
pion masses) considerably heavier than the values given experimentally
down to realistic values.  In particular, the use of a functional form
with $m_\pi$-dependence motivated by LDR ideas
was shown to lead to successful extraction of nucleon masses and magnetic
moments from lattice evaluations with $m_\pi$ in the 600 MeV range.

But isn't HB$\chi$pt independent of regularization scheme?  Yes, it is
true that any physical observable is independent of which scheme is
chosen, but a discussion of convergence involves more than just
the total observable.  The contribution to any observable from a
particular chiral order is not itself an observable, and yet the
comparison among chiral orders is what determines the rate of
convergence in HB$\chi$pt.  In dim reg, parameters in the Lagrangian
acquire a dependence on the regularization scale which exactly
cancels the scale dependence in the loops, thus making observables
independent of the unphysical dim reg scale.  LDR works the same way,
using the cutoff instead of the dim reg scale.  However an important
advantage of LDR, in contrast to dim reg, is that it does not ignore
loop contributions that contain positive powers of the cutoff.  The
Lagrangian parameters which exactly cancel these contributions are
not all at the same chiral order, so different numerical choices
for the cutoff change the relative sizes of the various orders in
the chiral expansion of an observable.  Since dim reg keeps only
the logarithmic divergences the relative sizes of chiral orders are
fixed in that scheme, and the inherent choice made by the dim reg
method unfortunately betrays a poor convergence in many observables
as has already been noted.  The present work indicates that, when
the LDR cutoff is fixed to its appropriate phenomenological value,
the LDR chiral expansion is superior.

In order to extend this work via application of other techniques such as
$1/N_c$ methods, it is also important to extend such calculations by 
inclusion of the decuplet intermediate state, since decuplet and octet
states become degenerate in this limit.  Indeed, since 
the mass difference between the nucleon and decuplet vanishes it is no
longer clear that the lowest excited state is that of the nucleon plus
Goldstone boson.  The interplay between the
chiral and $1/N_c$ limits is thus a subtle and interesting one\cite{coh}, 
and it is imperative to extend our previous LDR methods to include 
the case of the decuplet.  In order to be specific, we shall examine 
the same four problems (outlined above)
that were considered in our previous discussion and demonstrate that inclusion
of decuplet states can be handled in a parallel fashion to that done with
the octet.  In particular, we will see that any power divergences (possibly
accompanied by logarithms) involving
the cutoff can be absorbed into phenomenologically-determined constants so that
the remaining chiral corrections are small and do not destroy the traditional
$SU(3)$ fits to these processes.

In the next section then we show how to perform the dipole integrals 
relevant to decuplet inclusion and demonstrate that they have the expected
behavior in the chiral ($m_P<<\Lambda$) and heavy quark ($m_P>>\Lambda$)
limits.  In section III we 
apply these results to the specific problems discussed above, and in 
section IV we draw conclusions and set directions for future work.

\section{Integrals}
In this section we will review the decuplet integrals both in dim 
reg and in LDR. We begin with the integral which
appears in the analysis of the baryon masses
\begin{equation}   \label{massint}
\int\frac{{\rm d}^dk}{(2\pi)^d}\frac{k_ik_j}{(k_0-\Delta+i\epsilon)
(k^2-m^2+i\epsilon)} = -i\delta_{ij}\frac{I(m,\Delta)}{4 (d-1)\pi^2} .
\end{equation}
The dim reg result is
\begin{equation} \label{massdim}
I(m,\Delta) = \frac{\Delta}{2} \left(\frac{3}{2}m^2-\Delta^2\right)
  \left[ 2 \tilde{L} +  \ln \frac{m^2}{\mu^2} \right]  
   - (m^2-\Delta^2) \left(\frac{\Delta}{2}+ G(m,\Delta)\right)
\end{equation}
with 
\begin{equation}
G(m,\Delta) =
\left\{ \begin{array}{ll} 
                       (\Delta^2-m^2)^{1/2}\ln\left(\frac{\Delta}{m}
                      -\sqrt{\frac{\Delta^2}{m^2}-1}\right), & {\rm for}~
                      \Delta>m \\
                       (m^2-\Delta^2)^{1/2}\arccos\left(\frac{\Delta}{m}
                      \right), & {\rm for}~\Delta<m
                      \end{array}
              \right. 
\end{equation}
and
\begin{equation}
 \tilde{L} = \mu^{d-4} \left(\frac{1}{d-4}- \frac{1}{2}\big[\ln(4\pi)+1-\gamma\big] \right),
\end{equation}
where $\mu$ is the regularization scale.
The average octet-decuplet mass splitting $\Delta$ has a value of $\Delta = 231$
MeV and does not vanish in the chiral limit of vanishing quark masses.
The appearance of the mass scale $\Delta$ destroys the strict chiral 
counting scheme also in dim reg which has already been spoilt in the cutoff scheme
by introducing the scale $\Lambda$.
Since this splitting is only slightly larger than the pion mass but considerably
smaller than the kaon and eta masses, one expects the excitations of the decuplet 
to play an important role in $SU(3)$ $HB\chi$pt. 

The integral contains a divergent 
piece $\tilde{L}$ which has both mass-dependent and -independent pieces. The latter ones are
removed by redefining the common octet mass in the chiral limit, while the 
mass-dependent divergences can be absorbed into explicitly chiral symmetry breaking
counterterms at second chiral order, cf.~Ref.~\cite{bkm}. After renormalization the final
result is finite and any dependence on the regularization scale $\mu$ as given
by the logarithm of Eq.~(\ref{massdim}) is compensated by counterterms of
the same chiral order. This reflects the fact that chiral symmetry is preserved
by dim reg.
Albeit dim reg maintains chiral invariance, it fails in
separating the short and long distance components of the integral as was illustrated
in Ref.~\cite{ldr}. We would expect, {\it e.g.}, the long distance portion of the integral
$I(m,\Delta)$ to be smaller for larger meson masses. However, in the limit of
a large mass $m$ compared to $\Delta$ and the cutoff $\Lambda$,
\begin{equation}
I (m,\Delta) \stackrel{m \gg \Lambda, \Delta}{\longrightarrow} \; - \frac{\pi}{2} m^3 + \ldots,
\end{equation}
the function $I (m,\Delta)$ is found to have an $m^3$ dependence which implies
that the pion will contribute much less than its heavier kaon and eta counterparts
and is therefore in contradistinction to our intuitive expectation. An implicit
short distance contribution is carried along also if dim reg
is employed.

In Ref.~\cite{ldr} it was demonstrated that the long distance component of the integrals
is isolated --- and chiral invariance is simultaneously preserved --- by using
LDR.
Both an exponential cutoff and a simple dipole regulator 
$\Lambda^4/(\Lambda^2-k^2)^2$ were employed.
The latter one is more convenient for our purposes, since it enables the loop
integration to be carried out in simple analytic form. However, the specific shape 
of the cutoff is irrelevant as long as it is chirally invariant --- a consistent
chiral expansion can be carried out to the order we are working.

The introduction of a dipole cutoff in Eq.~(\ref{massint}) yields\cite{sig}
\begin{equation}   \label{massintcut}
\int\frac{{\rm d}^4k}{(2\pi)^4}\frac{k_ik_j}{(k_0-\Delta+i\epsilon)
(k^2-m^2+i\epsilon)} \left(  \frac{\Lambda^2}{\Lambda^2-k^2} \right)^2  
   = -i\delta_{ij}\frac{I_\Lambda(m,\Delta)}{12\pi^2}
\end{equation}
with
\begin{eqnarray}
I_\Lambda(m,\Delta) &=& \frac{\Lambda^4}{(\Lambda^2-m^2)^2}\left[
                        \frac{\Delta}{4}(\Lambda^2-m^2)
                      + \Delta\left(\Delta^2-\frac{3}{2}m^2\right)\ln\left(
                        \frac{\Lambda}{m}\right) \right. \nonumber \\
                   && - \sqrt{\Lambda^2-\Delta^2}\left(\frac{\Lambda^2}{2}
                        +\Delta^2-\frac{3m^2}{2}\right)
                        \arccos\left(\frac{\Delta}{\Lambda}\right)\nonumber \\
            && - \left.(m^2-\Delta^2)  G(m,\Delta)\right].
\end{eqnarray}
This time we obtain in the large $m$ limit
\begin{equation}
I_\Lambda (m,\Delta) \stackrel{m \gg \Lambda, \Delta}{\longrightarrow} \; 
   - \frac{\pi}{2} \frac{\Lambda^4}{m} + \ldots
\end{equation}
which is in conformity with our expectations since now the contributions from pions
are more important than those from kaons or etas.
LDR seems to be suited to disentangle the long distance
part of the integral we are interested in from the spurious high energy portion.
As we will see in the next section it also preserves chiral invariance for
the cases considered here and provides an alternative regularization scheme
for HB$\chi$pt.

It is instructive to consider the opposite limit of the integral $I_\Lambda$:
for large values of the cutoff we obtain
\begin{eqnarray}  \label{smallmass}
I_\Lambda(m,\Delta) &=& -\frac{\pi\Lambda^3}{4} + \frac{3}{4}\Delta\Lambda^2
                      - \frac{\pi}{8}\Lambda(3\Delta^2-2m^2)
                      + \frac{5}{6}\Delta^3 - \frac{m^2\Delta}{4}
                      \nonumber \\
                   && + \Delta\left(\Delta^2-\frac{3}{2}m^2\right)\ln\left(
                        \frac{\Lambda}{m}\right) - (m^2-\Delta^2)  G(m,\Delta)\nonumber \\
               && + O(1/\Lambda) .
\end{eqnarray}
One observes cubic, quadratic, linear and logarithmic divergences which can
be absorbed by appropriate counterterms, see Sec. \ref{sec:cont}. 
The nonanalytic terms in the quark
masses, on the other hand, reduce to the dim reg result.

We now turn to the Feynman integrals of the remaining cases.
In the analysis of S-wave hyperon decays
the relevant heavy baryon integral is
\begin{equation}   \label{axint}
\int\frac{{\rm d}^dk}{(2\pi)^d}\frac{k_ik_j}{(k_0-\Delta+i\epsilon)^2
(k^2-m^2+i\epsilon)} = -i\delta_{ij}\frac{3 J(m,\Delta)}{16 (d-1)\pi^2} .
\end{equation}
In dim reg we obtain
\begin{equation} \label{axdim}
J(m,\Delta) =  \left(m^2- 2\Delta^2\right)
  \left[ 2 \tilde{L} +  \ln \frac{m^2}{\mu^2} \right]  + \frac{2}{3}\left(m^2 
          + \Delta^2 \right)  + 4 \Delta G(m,\Delta) ,     
\end{equation}
whereas the LDR version is given by
\begin{equation}   
\int\frac{{\rm d}^4k}{(2\pi)^4}\frac{k_ik_j}{(k_0-\Delta+i\epsilon)^2
(k^2-m^2+i\epsilon)} \left(  \frac{\Lambda^2}{\Lambda^2-k^2} \right)^2
= -i\delta_{ij}\frac{ J_\Lambda(m,\Delta)}{16 \pi^2}
\end{equation}
with
\begin{eqnarray}
J_\Lambda(m,\Delta) &=& \frac{\Lambda^4}{(\Lambda^2-m^2)^2}\left[
                        \Lambda^2-m^2
                      + \left(m^2-2\Delta^2 \right)\ln\left(
                        \frac{m^2}{\Lambda^2}\right) \right. \nonumber \\
                   & & - \frac{2 \Delta}{\sqrt{\Lambda^2-\Delta^2}}\left(\Lambda^2
                        +m^2- 2\Delta^2\right)
                        \arccos\left(\frac{\Delta}{\Lambda}\right) \nonumber \\
                   & &\left. + 4 \Delta G(m,\Delta)\right]  .
\end{eqnarray}
In the limit of a large cutoff the integral simplifies to
\begin{eqnarray} \label{swaexp}
J_\Lambda(m,\Delta) &\stackrel{m , \Delta \ll \Lambda}{\longrightarrow} & \;
              \Lambda^2 + m^2 + [m^2 - 2\Delta^2] \ln\left(
                        \frac{m^2}{\Lambda^2}\right) 
                      - \pi \Delta \Lambda + 2 \Delta^2 
                      \nonumber \\
                   & &  + 4 \Delta G(m,\Delta) \quad + O(1/\Lambda) 
\end{eqnarray}
which are the nonanalytic pieces of the dim reg result
plus a polynomial in $m^2$. 
In the opposite limit of large meson masses we obtain
\begin{eqnarray}
J_\Lambda(m,\Delta) & \stackrel{m \gg  \Lambda, \Delta}{\longrightarrow} & \;
                \frac{\Lambda^4}{m^2}\left[
                        -1
                 + \ln\left(\frac{m^2}{\Lambda^2}\right)
                   - \frac{2 \Delta}{\sqrt{\Lambda^2-\Delta^2}}
                        \arccos\left(\frac{\Delta}{\Lambda}\right)\right]
\end{eqnarray}
and again our intuitive expectations are met.

In the calculation of the baryon axial couplings another integral enters
in addition to Eq.~(\ref{axint}), cf. Ref.~\cite{bac},
\begin{equation}   \label{axint2}
\int\frac{{\rm d}^dk}{(2\pi)^d}\frac{k_ik_j}{(k_0-\Delta+i\epsilon)(k_0+i\epsilon)
(k^2-m^2+i\epsilon)} = -i\delta_{ij}\frac{ \bar{J}(m,\Delta)}{4 (d-1)\pi^2} .
\end{equation}
In dim reg the integral reads
\begin{eqnarray}
\bar{J}(m,\Delta) &=& \frac{1}{\Delta} \left[I(m,\Delta) - I(m,0)\right] \nonumber\\
    &=& \frac{1}{2} \left(\frac{3}{2}m^2-\Delta^2\right)
  \left[ 2 \tilde{L} +  \ln \frac{m^2}{\mu^2} \right]  \nonumber\\
 &&   - (m^2-\Delta^2) \left(\frac{1}{2}+ \frac{1}{\Delta}G(m,\Delta)\right) 
 + \frac{\pi}{2 } \frac{m^3}{\Delta} .
\end{eqnarray}
For the LDR version,
\begin{equation}   
\int\frac{{\rm d}^4k}{(2\pi)^4}\frac{k_ik_j}{(k_0-\Delta+i\epsilon)(k_0+i\epsilon)
(k^2-m^2+i\epsilon)} \left(  \frac{\Lambda^2}{\Lambda^2-k^2} \right)^2
\!  = -i\delta_{ij}\frac{ \bar{J}_\Lambda(m,\Delta)}{12 \pi^2},
\end{equation}
we obtain
\begin{eqnarray}
\bar{J}_\Lambda(m,\Delta) &=& \frac{\Lambda^4}{(\Lambda^2-m^2)^2}\left[
                        \frac{1}{4}(\Lambda^2-m^2)
                      + \left(\Delta^2-\frac{3}{2}m^2\right)\ln\left(
                        \frac{\Lambda}{m}\right) \right. \nonumber \\
                   && - \sqrt{\frac{\Lambda^2}{\Delta^2}-1}\left(\frac{\Lambda^2}{2}
                        +\Delta^2-\frac{3m^2}{2}\right)
                        \arccos\left(\frac{\Delta}{\Lambda}\right)\nonumber \\
            && - \left.\frac{1}{\Delta}(m^2-\Delta^2)  G(m,\Delta) + \frac{\pi}{4\Delta} \Big(
                   \Lambda^3 - 3 m^2 \Lambda +2 m^3 \Big) \right] . 
\end{eqnarray}
This yields, in the limit $\Lambda \rightarrow \infty$,
\begin{eqnarray}   \label{axexp}
\bar{J}_\Lambda(m,\Delta) &=&  \frac{3}{4} \Lambda^2
                      - \frac{3\pi}{8}\Lambda \Delta
                      + \frac{5}{6}\Delta^2 - \frac{m^2}{4} + \frac{\pi}{2} \frac{m^3}{\Delta}
                      \nonumber \\
                   && + \left(\Delta^2-\frac{3}{2}m^2\right)\ln\left(
                        \frac{\Lambda}{m}\right) - \frac{1}{\Delta}(m^2-\Delta^2)  G(m,\Delta)\nonumber \\
               && + O(1/\Lambda),
\end{eqnarray}
and for large meson masses
\begin{equation}
\bar{J}_\Lambda(m,\Delta)  \stackrel{m \gg  \Lambda, \Delta}{\longrightarrow}  \;
                \frac{3\Lambda^4}{4m^2}\left[
                        1  + \ln\left(\frac{m^2}{\Lambda^2}\right)
               + 2 \sqrt{\frac{\Lambda^2}{\Delta^2}-1}
             \arccos\left(\frac{\Delta}{\Lambda}\right) - \pi \frac{\Lambda}{\Delta}\right].
\end{equation}

Finally, we consider the integral involved in the calculation of the 
magnetic moments. It reads
\begin{equation}   \label{magint}
\int\frac{{\rm d}^dk}{(2\pi)^d}\frac{k_ik_j}{(k_0-\Delta+i\epsilon)
(k^2-m^2+i\epsilon)^2} = -i\delta_{ij}\frac{3 K(m,\Delta)}{16 (d-1)\pi^2}
\end{equation}
where
\begin{equation} \label{magdim}
K(m,\Delta) = \Delta \left[ 2 \tilde{L} +  \ln \frac{m^2}{\mu^2} \right]  
            - \frac{\Delta}{3}  -2 G(m,\Delta) .
\end{equation}
In LDR one obtains
\begin{equation}   
\int\frac{{\rm d}^4k}{(2\pi)^4}\frac{k_ik_j}{(k_0-\Delta+i\epsilon)
(k^2-m^2+i\epsilon)^2} \left(  \frac{\Lambda^2}{\Lambda^2-k^2} \right)^2
 = -i\delta_{ij}\frac{ K_\Lambda(m,\Delta)}{16 \pi^2}
\end{equation}
with 
\begin{eqnarray}
K_\Lambda(m,\Delta) &=& \frac{\Lambda^4}{(\Lambda^2-m^2)^3}\left[
                        \frac{2}{3} \Delta [\Lambda^2-m^2]
                      + \Delta \left[m^2+ \Lambda^2-\frac{4}{3}\Delta^2 \right]\ln\left(
                        \frac{m^2}{\Lambda^2}\right) \right. \nonumber \\
                   & & + 2 \left[m^2+ \frac{1}{3}\Lambda^2-\frac{4}{3}\Delta^2 \right] 
                   {\sqrt{\Lambda^2-\Delta^2}}
                        \arccos\left(\frac{\Delta}{\Lambda}\right)\nonumber \\
            && - \left.\frac{2}{3} \left[m^2+ 3 \Lambda^2-4\Delta^2 \right] G(m,\Delta)\right].
\end{eqnarray}
The small and large mass limits of the LDR form are given by
\begin{equation}
K_\Lambda(m,\Delta) \stackrel{m , \Delta \ll \Lambda}{\longrightarrow}  \;
              \Delta  \ln \left( \frac{m^2}{\Lambda^2} \right) 
            + \frac{\pi}{3} \Lambda + \frac{2}{3} \Delta -2 G(m,\Delta) \quad + O(1/\Lambda) 
\end{equation}
and
\begin{equation}
K_\Lambda(m,\Delta)  \stackrel{m \gg  \Lambda, \Delta}{\longrightarrow}  \;
                \frac{\pi}{3} \frac{\Lambda^4}{m^3} + \ldots
\end{equation}
This discussion of the relevant integrals in the analysis of the baryon
masses, axial couplings, S-wave hyperon decays, and magnetic moments has
revealed that LDR is well suited to separating the long distance physics
from the high energy portion of the integrals, whereas dim reg
includes large contributions from short distance physics. With LDR as
a regularization technique for handling loop integrals, we expect to ameliorate
problems which have arisen in previous calculations where dim reg produced
large loop effects.

\section{Decuplet contributions} \label{sec:cont}
We now turn to the calculation of the decuplet contributions in the 
four processes discussed above and give numerical results.
In the present investigation we will restrict ourselves to the presentation
of loop integrals with an internal decuplet field. (For an evaluation of the
octet baryon loops the reader is referred to Ref.~\cite{ldr}.)

Up to linear order in the derivative expansion the effective Lagrangian 
in the heavy baryon formulation is given by
\begin{eqnarray} \label{declag}
{\cal L}_{\phi B T} &=& i \langle \bar{B} [v \cdot D, B] \rangle
                     + D  \langle \bar{B} S_\mu\{u^\mu, B\} \rangle
                     + F  \langle \bar{B} S_\mu [u^\mu, B] \rangle\nonumber \\
                 && 
 -  i \bar{T}_\mu v \cdot D T^\mu 
                   + \Delta \bar{T}_\mu T^\mu + \frac{{\cal C}}{2} \left( 
                    \bar{T}_\mu u^\mu B + \bar{B}u^\mu T_\mu \right) \nonumber \\
                 &&  + H  \bar{T}_\mu S_\nu u^\nu  T^\mu + h_c \bar{T}_\mu h_+  T^\mu,
\end{eqnarray}
where $\langle \ldots \rangle$ denotes the trace in flavor space and we have suppressed
flavor indices for the decuplet terms.
The first three terms contain the kinetic parts of the octet baryon fields $B = (N,
\Lambda, \Sigma, \Xi)$\footnote{We work in the isospin limit of equal
up and down quark masses.} and the axial-vector coupling of the mesons to the octet 
baryons. The covariant derivative $D_\mu$ for the baryon fields enters in 
combination with the four-velocity $v_\mu$ and 
$2 S_\mu = i \gamma_5 \sigma_{\mu \nu} v^\nu$ is
the Pauli-Lubanski spin vector.
For the coupling constants $D$ and $F$ we use the values in the $SU(6)$ limit,
$D= 0.75 , F = 0.5 $, which yield $g_A = 1.25$.
The Goldstone bosons are summarized in
\begin{equation}
u_\mu = i u^\dagger \nabla_\mu U u^\dagger, \qquad
U= u^2 = \exp \Big( \frac{i}{F_\pi} \sum_j \lambda_j \phi_j \Big),
\end{equation}
with $\nabla_\mu $ being the covariant derivative of the meson fields
and $F_\pi = 92.4$ MeV the pion decay constant.
A mass term is still present for the decuplet fields 
$T = (\Delta, \Sigma^*, \Xi^*, \Omega)$
after transforming to the heavy mass formulation due the
octet-decuplet mass splitting $\Delta$ which does not vanish in the chiral limit.
In the Feynman rules the mass splitting $\Delta$ is contained in the 
decuplet propagator
\begin{equation}
\frac{i}{ v \cdot l \, \, - \Delta + \, \, i \epsilon}      
\, \, \bigg( v_{\mu} v_{\nu} - g_{\mu \nu} 
-\frac{4}{d-1}S_{\mu}S_{\nu} \bigg)
\end{equation}
in $d$ dimensions.
The appearance of the mass scale $\Delta$ destroys in the case of
dim reg the one--to--one correspondence between
meson loops and the expansion in small momenta and quark masses.
No further complications arise in our case since the strict chiral
counting scheme has already been spoilt by introducing the scale $\Lambda$.

The sixth term in Eq.~(\ref{declag}) is responsible for strong decays of the decuplet into a ground
state baryon and a pseudoscalar meson. The pertinent coupling constant 
$|{\cal C}| = 1.5 \pm 0.3$ can be determined from the decay $\Delta \rightarrow N \pi$.
The seventh term describes the axial-vector coupling of the Goldstone bosons
to the decuplet fields, whereas the last term induces weak nonleptonic
decay of the decuplet fields. We have defined $h_+ = u^\dagger h u$ where
$h^a_b = \delta^a_2 \delta^3_b$ is the weak transition matrix.
The weak coupling $h_c$ has been determind from a fit to the nonleptonic
hyperon decays in Ref.~\cite{jen}, $h_c = (-0.10 \pm  0.19) \times 10^{-7}$ GeV. 
Although in that work only the leading nonanalytic terms 
of the chiral loops were taken into account using dim reg, 
the light quark masses
$m_u, m_d$ were set to zero, and the octet-decuplet mass splitting was neglected,
we will use the resulting value for $h_c$ due to the lack of any more general
study in the literature.
The parameter $H$ can, in principle, be determined from a fit to the baryon
axial currents as has been done in Ref.~\cite{bac} up to one-loop order. However, 
the value for $H$ depends strongly on the chosen cutoff scale and differs 
considerably  from the value in dim reg. One obtains
$H= 3.0 \pm 5.0$ in the cutoff scheme, whereas in the case of dim reg
a fit yields $H=-3.5$, and thus no reliable estimate for the parameter $H$
can be given in chiral perturbation theory.
An independent estimate for $H$ has recently been obtained in large $N_c$
QCD\cite{ram} where the authors obtain values in the range $H = -2.25 \ldots -0.66$
from performing different fits to the octet baryon axial currents.

\subsection{Baryon masses}\label{masscalc}
The decuplet contributions to the octet baryon masses in dim reg
involve the integral given in Eq.~(\ref{massint}),
\begin{equation}
\delta M_B = \frac{{\cal C}^2}{24\pi^2F_\pi^2}\sum_{\phi=\pi,K,\eta}
             \alpha_B^\phi I(m_\phi,\Delta),
\end{equation}
with the coefficients
\begin{equation}
\begin{array}{llllll}
\alpha_N^\pi=4, &\alpha_N^K=1, &\alpha_N^\eta=0 
&\alpha_\Sigma^\pi=2/3, &\alpha_\Sigma^K=10/3, &\alpha_\Sigma^\eta=1, \\[0.1cm]
\alpha_\Lambda^\pi=3, &\alpha_\Lambda^K=2, &\alpha_\Lambda^\eta=0 
&\alpha_\Xi^\pi=1, &\alpha_\Xi^K=3, &\alpha_\Xi^\eta=1 .
\end{array}
\end{equation}
By using the following inputs, $m_\pi = 138 {\rm MeV},  m_K = 495{\rm MeV}$
and $m_\eta = \sqrt{(4m_K^2-m_\pi^2)/3} = 566$ MeV which is the value from the Gell-Mann---Okubo
mass relation for the pseudoscalar mesons, $|{\cal C}| =1.5$, the
regularization scale 
$\mu = 1$ GeV, omitting the piece proportional to $\tilde{L}$  and setting $d=4$
we obtain the dim reg results in Table \ref{masstabdec} for the sum 
of octet and decuplet contributions.
Strictly speaking, the $d$-dependence of the decuplet propagator and the definition of the integral
$I$ in Eq. (\ref{massint}) will yield additional analytic contributions in the $d \rightarrow 4$
limit due to the divergent piece $\tilde{L}$. However, these are 
numerically unimportant and can always be removed by appropriate counterterms
so that our conclusions remain unchanged for the four processes discussed in this work.
In order to keep the formulae and the discussion as simple as possible, we therefore
drop the $\tilde{L}$ term and perform the limit $d \rightarrow 4$ afterwards.

For comparison we also present in Table \ref{masstaboct}  the contributions of
the octet loops only.
In both of the dim reg cases we observe large mass shifts which upset the convergence of the chiral series.

\begin{table}[th]  
\begin{center}\begin{tabular}{rrrrrr}  
      & Dim. & $\Lambda$=300 & $\Lambda$=400 & $\Lambda$=500 & $\Lambda$=600 \\
\hline
$N$       & -0.46 & -0.04 & -0.03 & -0.00 &  0.02 \\
$\Sigma$  & -1.48 & -0.02 &  0.02 &  0.07 &  0.13 \\
$\Lambda$ & -1.02 & -0.03 &  0.00 &  0.04 &  0.10 \\
$\Xi$     & -1.83 & -0.01 &  0.04 &  0.10 &  0.18
\end{tabular}
\caption{Sum of octet and decuplet contributions to the octet baryon masses in GeV both in dimensional
         regularization and for various values of the cutoff $\Lambda$ in MeV.}
\label{masstabdec}
\end{center}
\end{table}

\begin{table}[ht] 
\begin{center}\begin{tabular}{rrrrrr}        
       & Dim. & $\Lambda$=300 & $\Lambda$=400 & $\Lambda$=500 & $\Lambda$=600 \\
\hline
$N$       & -0.31 &  0.02 &  0.03 &  0.05 &  0.07 \\
$\Sigma$  & -0.62 &  0.03 &  0.05 &  0.08 &  0.12 \\
$\Lambda$ & -0.69 &  0.03 &  0.06 &  0.09 &  0.13 \\
$\Xi$     & -1.03 &  0.04 &  0.08 &  0.12 &  0.17
\end{tabular}
\caption{Octet contributions to the octet baryon masses in GeV both in dimensional
         regularization and for various values of the cutoff $\Lambda$ in MeV.}
\label{masstaboct}
\end{center}
\end{table}

We now repeat the calculation in the LDR scheme by employing the integral
from Eq.~(\ref{massintcut}) for the decuplet contributions.\footnote{For the explicit
expressions for the octet loop contributions see Ref.~\cite{ldr} and also 
Ref.~\cite{sig}.} 
Before producing numerical results, however,
it is necessary to confirm that LDR preserves chiral invariance
and is suited for evaluating loop integrations in HB$\chi$pt. To this end, we utilize
the large cutoff expansion of the integral in Eq.~(\ref{smallmass}),

\begin{eqnarray}  \label{massexp}
\delta M_B &=& \frac{{\cal C}^2}{24\pi^2F_\pi^2}\sum_{\phi=\pi,K,\eta}
             \alpha_B^\phi \left(   -\frac{\pi\Lambda^3}{4} + \frac{3}{4}\Delta\Lambda^2
                      - \frac{\pi}{8}\Lambda(3\Delta^2-2m_\phi^2) \right. \nonumber \\
                &&  \left. + \Delta\left(\Delta^2-\frac{3}{2}m_\phi^2\right)\ln \Lambda 
                + \ldots \right),
\end{eqnarray}
where the ellipsis denotes terms which remain finite as $\Lambda \rightarrow \infty$.
The quark mass independent pieces can be absorbed into a renormalization of the
octet baryon mass in the chiral limit, $M_0$, by defining
\begin{equation}
M_0^r = M_0 + \frac{5 {\cal C}^2}{24\pi^2F_\pi^2} \Big[
              -\frac{\pi}{4}\Lambda^3 + \frac{3}{4}\Delta\Lambda^2
             - \frac{3 \pi}{8}\Lambda \Delta^2  + \Delta^3\ln \Lambda + \pi \Delta^3 \Big] .
\end{equation}   
Note that we have also absorbed a constant term $\pi \Delta^3$ which
does not depend on the cutoff scale $\Lambda$ for reasons to be explained after 
Eq.~(\ref{massren}).   
The renormalization of $M_0$ already occurs in dim reg
since the introduction of the mass scale $\Delta$ spoils the strict chiral
power counting\cite{bkm}.
In order to absorb the mass-dependent components in Eq.~(\ref{massexp}),
one must introduce explicitly chiral symmetry breaking counterterms which are
proportional to the quark masses.
They are given by
\begin{equation}
{\cal L}_\chi =  b_D \langle \bar{B}  \{\chi_+,B\} \rangle +
                     b_F \langle \bar{B}  [\chi_+,B] \rangle 
                     + b_0 \langle \bar{B}B \rangle \langle \chi_+ \rangle .
\end{equation}
The quark mass matrix ${\cal M}
= \mbox{diag}(m_u, m_d,m_s)   $
enters in the combination $\chi_+ = 2 B_0 (u^\dagger {\cal M} u^\dagger + u {\cal M} u  )$
with $B_0 = - \langle  0 | \bar{q} q | 0\rangle/ F_\pi^2$ the order
parameter of the spontaneous symmetry violation.
The low-energy constants $b_D$ and $b_F$ are responsible for the splitting of the baryon octet masses 
at leading order in symmetry breaking.
By redefining
\begin{equation}
b_D^r = b_D + \frac{1}{2} c , \quad
b_F^r = b_F - \frac{5}{12} c , \quad
b_0^r = b_0 - \frac{7}{6}  c 
\end{equation}
with 
\begin{equation}
c = \frac{{\cal C}^2}{24 \pi^2 F_\pi^2} \left( \frac{ \pi}{4} \Lambda 
 - \frac{3}{2} \Delta  \ln \Lambda   \right).
\end{equation}
one can absorb the mass-dependent pieces in Eq.~(\ref{massexp}) into the low-energy
constants. This proves the chiral invariance of the cutoff procedure and shows
that the final physics is independent of $\Lambda$ if performed to all orders
since one can remove the 
$\Lambda$ dependence by renormalizing the couplings of the counterterms.

Having convinced ourselves from the chiral invariance of the cutoff procedure
we can now vary the cutoff in the phenomenologically relevant range
300 MeV $<\Lambda<$ 600 MeV. But we first remove the asymptotic 
mass-independent components of the function $I_\Lambda$ by defining
\begin{equation} \label{massren}
I^r_\Lambda = I_\Lambda +\frac{\pi}{4}\Lambda^3 - \frac{3}{4}\Delta\Lambda^2
+\frac{3 \pi}{8}\Lambda \Delta^2  - \Delta^3\ln \Lambda -  \pi \Delta^3 
\end{equation}
since these effects can be absorbed into $M_0$ and yield misleading indications
about the size of the nonanalytic effects in the large cutoff limit.
As mentioned before, we have also subtracted a constant term $ \pi \Delta^3$
which ensures that the original integral $I$ is not altered much numerically
for physically relevant values of the cutoff, 300 MeV $\le \Lambda \le$ 600 MeV.
Its exact value is not of importance, it has just been chosen in such a way that it compensates to a large
extent the contribution from the power and log divergences in Eq.~(\ref{massren});
in fact, for $\Lambda \simeq 450$ MeV one obtains $I^r_\Lambda \approx I_\Lambda$ in Eq.~(\ref{massren}).
This renormalization procedure allows one both to reduce the $\Lambda$-dependence
of the integrals even for larger values of the cutoff $\Lambda$ and to conserve
approximately the numerical value of the original integral.

The long distance decuplet contribution to the octet baryon masses is then
given by
\begin{equation}
\delta M_B = \frac{{\cal C}^2}{24\pi^2F_\pi^2}\sum_{\phi=\pi,K,\eta}
             \alpha_B^\phi I^r_\Lambda(m_\phi,\Delta)
\end{equation}
and the corresponding numerical results are shown  in Tables \ref{masstabdec}
and \ref{masstaboct}.
Clearly, the convergence of the chiral series seems to be under control.
Moreover, what matters for $SU(3)$ breaking are differences in the loop 
contributions to the different octet baryon masses, since a constant effect
can be absorbed into the common octet mass $M_0$. This difference is quite
small for the cutoff version, whereas the $SU(3)$ breaking effects are significantly
larger in dim reg due to the inclusion of spurious
short distance physics. 

It is also important to consider the quality of the HB$\chi$pt fit to
experimental data.  To accomplish this, we combine the experimental
uncertainty of each observable with an overall theoretical uncertainty
arising from our neglect of $O(p^4)$ terms (and beyond) in HB$\chi$pt.
In particular we determine the theoretical uncertainty from the assumption
that the expansion parameter of HB$\chi$pt is 30\%, which is between
$m_\pi/m_N$ and $m_K/m_\Sigma$.  If our assumption is valid, then our fits
will have $\chi^2/d.o.f.$ of order unity or smaller and the HB$\chi$pt
fits will then be considered successful.

In the case of the baryon masses, which begin at $O(1)$ in HB$\chi$pt,
our theoretical uncertainty is (1GeV)$\times(30\%)^4$, and a very good fit
can be found both in lowest order and with chiral loop corrections.
The former is simply the traditional $SU(3)$ fit and yields a 
$\chi^2/d.o.f.=0.61$ using the parameters
$$b_D=0.066\,\,{\rm GeV}^{-1},\,\,\,b_F=-0.209\,\,{\rm
GeV}^{-1},\,\,\,\bar{M}_0=1.20\,\,{\rm GeV},$$
while with the inclusion of octet plus decuplet (octet only) loops we find 
the results given in Table 3 (Table 4) respectively.
Notice that all of these mass fits satisfy our success criterion.

\begin{table}
\begin{tabular}{cccccc}

 &Dim.&$\Lambda$=300&$\Lambda$=400&$\Lambda=500$&$\Lambda$=600\\
\hline
$b_D[{\rm GeV}^{-1}]$&0.447&0.061&0.055&0.047&0.036\\
$b_F[{\rm GeV}^{-1}]$&-0.97&-0.19&-0.17&-0.15&-0.12\\
$\bar{M}_0$[GeV]&2.71&1.22&1.18&1.13&1.06\\
$\chi^2/d.o.f.$&0.24&0.56&0.51&0.46&0.42
\end{tabular}
\caption{Baryon mass fits with octet plus decuplet loop contributions and
calculated with meson couplings having their $SU(6)$ values --- $D$=0.75,
$F$=0.50, $C=-1.5$.}
\end{table} 

\begin{table}
\begin{tabular}{cccccc}

 &Dim.&$\Lambda$=300&$\Lambda$=400&$\Lambda=500$&$\Lambda$=600\\
\hline
$b_D[{\rm GeV}^{-1}]$&0.008&0.068&0.070&0.072&0.075\\
$b_F[{\rm GeV}^{-1}]$&-0.61&-0.20&-0.19&-0.17&-0.15\\
$\bar{M}_0$[GeV]&1.81&1.17&1.14&1.11&1.08\\
$\chi^2/d.o.f.$&0.07&0.58&0.56&0.53&0.51
\end{tabular}
\caption{Baryon mass fits with octet only loop contributions and
calculated with meson couplings having their $SU(6)$ values --- $D$=0.75, $F$=0.50.}
\end{table}

\subsection{Axial couplings}
Next, we consider the decuplet contributions to the octet baryon axial couplings
which have the form
\begin{eqnarray}
g_A [ij]  &=& - \frac{C^2}{24 \pi^2 F_\pi^2}\:
      \sum_{\phi=\pi,K,\eta} \: \bigg(  -  \frac{4}{3} 
         \rho_{ij}^{\, \phi} \bar{J}(m_\phi, \Delta)\nonumber \\
  &&     +\frac{5}{6}  H 
        \:\sigma_{ij}^{\, \phi} J(m_\phi, \Delta)   + \frac
  {3}{2} \gamma_{ij} (\epsilon_i^{\, \phi}  + \epsilon_j^{\, \phi} )
      J(m_\phi, \Delta)     \bigg) \; . 
\end{eqnarray}
The last term with the coefficients 
$\epsilon_{i}^{\, \phi}$ takes into account the contributions
from intermediate decuplet states to the octet baryon $Z$--factors.
The coefficients read

\begin{equation}
\begin{array}{lll}
\rho_{pn}^{\pi}    =   \frac{8}{3} ( D+F)  , 
&\rho_{pn}^K    =    D + \frac{1}{3} F ,
&\rho_{pn}^{\eta}    =   0  ; \\[0.3cm]
\rho_{\Lambda \Sigma^-}^{\pi}    =    \frac{1}{3\sqrt{6}} ( D+6F) ,
&\rho_{\Lambda \Sigma^-}^K  = \frac{1}{\sqrt{6}} (\frac{8}{3} D + 4F) ,
&\rho_{\Lambda \Sigma^-}^{\eta}    =  \frac{1}{\sqrt{6}} D ; \\[0.3cm]
\rho_{\Xi^0 \Xi^-}^{\pi}    =   - \frac{1}{3} ( D-F) , 
&\rho_{\Xi^0 \Xi^-}^K    =   \frac{1}{3} (D+5F) ,
&\rho_{\Xi^0 \Xi^-}^{\eta}    =   \frac{1}{3} (D+3F)  ; \\[0.3cm]
\rho_{p \Lambda}^{\pi}    =   - \frac{3}{2\sqrt{6}} (\frac{11}{3}D+F) ,
&\rho_{p \Lambda}^K    =  - \frac{3}{2\sqrt{6}} (D+F) ,
&\rho_{p \Lambda}^{\eta}    =   0  ; \\[0.3cm]
\rho_{\Lambda \Xi^- }^{\pi}    =    \frac{3}{2 \sqrt{6}} (\frac{1}{3}D-F) , 
&\rho_{\Lambda \Xi^- }^K  = \frac{3}{2 \sqrt{6}} (D-F)  ,
&\rho_{\Lambda \Xi^- }^{\eta}    =  \frac{1}{\sqrt{6}} D \; \\[0.3cm]
\rho_{n \Sigma^-}^{\pi}    =   \frac{1}{3} (D+5F) , 
&\rho_{n \Sigma^-}^K    =    \frac{1}{6} (D+5F) ,
&\rho_{n \Sigma^-}^{\eta}    =   - \frac{1}{6} (D-3F)  ; \\[0.3cm]
\rho_{\Sigma^0 \Xi^-}^{\pi}    =    \frac{1}{3 \sqrt{2}} ( 2D+F) ,
&\rho_{\Sigma^0 \Xi^-}^K  = \frac{1}{6\sqrt{2}} (15D+13F) ,
&\rho_{\Sigma^0 \Xi^-}^{\eta}    =  \frac{1}{2\sqrt{2}} (D+F) ; \\[0.3cm]
\rho_{\Sigma^+ \Xi^0}^{\phi}  =  \sqrt{2}\rho_{\Sigma^0 \Xi^-}^{\phi}. &&
\end{array}
\end{equation}

\begin{equation}
\begin{array}{lll}
\sigma_{pn}^{\pi} = 10/9 ,
&\sigma_{pn}^{K} = 2/9  ,
&\sigma_{pn}^{\eta} = 0  , \\[0.3cm]
\sigma_{\Lambda \Sigma^-}^{\pi} = 2/(3 \sqrt{6})  ,
&\sigma_{\Lambda \Sigma^-}^{K} =   1/(3 \sqrt{6}) ,
&\sigma_{\Lambda \Sigma^-}^{\eta} = 0  , \\[0.3cm]
\sigma_{\Xi^0 \Xi^-}^{\pi} =1/18  ,
&\sigma_{\Xi^0 \Xi^-}^{K} = - 2/9  ,
&\sigma_{\Xi^0 \Xi^-}^{\eta} = - 1/6  , \\[0.3cm]
\sigma_{p \Lambda}^{\pi} = - 2/\sqrt{6}  ,
&\sigma_{p \Lambda}^{K} = - 1/\sqrt{6}  ,
&\sigma_{p \Lambda}^{\eta} = 0 , \\[0.3cm]
\sigma_{\Lambda \Xi^-}^{\pi} = 1/\sqrt{6}  ,
&\sigma_{\Lambda \Xi^-}^{K} = 1/\sqrt{6}  ,
&\sigma_{\Lambda \Xi^-}^{\eta} = 0  , \\[0.3cm]
\sigma_{n \Sigma^-}^{\pi} = - 2/9  ,
&\sigma_{n \Sigma^-}^{K} = - 1/9  ,
&\sigma_{n \Sigma^-}^{\eta} =  0  , \\[0.3cm]
\sigma_{\Sigma^0 \Xi^-}^{\pi} = 2/(9\sqrt{2})  ,
&\sigma_{\Sigma^0 \Xi^-}^{K} = 7/(9\sqrt{2})  ,
&\sigma_{\Sigma^0 \Xi^-}^{\eta} = 1/(3\sqrt{2})  , \\[0.3cm]
\sigma_{\Sigma^0 \Xi^-}^{\phi}  = \sqrt{2} \sigma_{\Sigma^0 \Xi^-}^{\phi}.&&
\end{array}
\end{equation}

\begin{equation}
\begin{array}{lll}
\gamma_{pn} = D+F ,   & \gamma_{\Lambda \Sigma^-} = 2D/\sqrt{6} ,
&\gamma_{\Xi^0 \Xi^-} = D-F , \\[0.3cm]
\gamma_{p\Lambda} = -(D+3F)/\sqrt{6} ,
&\gamma_{\Lambda \Xi^-} = -(D-3F)/\sqrt{6} , & \gamma_{n \Sigma^-} = D-F, 
\\[0.3cm]
\gamma_{\Sigma^0 \Xi^-} = (D+F)/\sqrt{2}, & \gamma_{\Sigma^+ \Xi^0} = D+F .&
\end{array}
\end{equation}

\begin{equation}
\begin{array}{llllll}
\epsilon_N^{\pi}  =  1  , 
&\epsilon_N^{K}  =  1/4 ,
&\epsilon_N^{\eta}  =  0  ,
&\epsilon_\Sigma^{\pi}  =  1/6,
&\epsilon_\Sigma^{K}  =  5/6 ,
&\epsilon_\Sigma^{\eta}  =  1/4   \\[0.3cm]
\epsilon_\Lambda^{\pi}  =  3/4 ,
&\epsilon_\Lambda^{K}  =  1/2 \,
&\epsilon_\Lambda^{\eta}  =  0   ,  
&\epsilon_\Xi^{\pi}  =  1/4 ,
&\epsilon_\Xi^{K}  =  3/4 ,
&\epsilon_\Xi^{\eta}  =  1/4 .
\end{array}
\end{equation}
The numerical results for the finite pieces can be found in Tables \ref{axtabdec}
and \ref{axtaboct}. We have used the value $H=-2.0$
which was obtained in Ref.~\cite{ram}.

\begin{table}[ht]
\begin{center}
\begin{tabular}{lrrrrr} 
      & Dim. & $\Lambda$=300 & $\Lambda$=400 & $\Lambda$=500 & $\Lambda$=600 \\
\hline
$g_A(\bar{p}n)$            &  0.63 &  0.00 &  0.10 &  0.19 &  0.27 \\
$g_A(\bar{p}\Lambda)$      & -0.83 & -0.06 & -0.15 & -0.24 & -0.32 \\
$g_A(\bar\Lambda\Sigma^-)$ &  0.29 & -0.03 &  0.03 &  0.07 &  0.12 \\
$g_A(\bar{n}\Sigma^-)$     & -0.14 & -0.08 & -0.05 & -0.02 & -0.01 \\
$g_A(\bar\Lambda\Xi^-)$    &  0.57 &  0.09 &  0.11 &  0.14 &  0.16 \\
$g_A(\bar\Sigma^0\Xi^-)$   &  0.83 &  0.03 &  0.13 &  0.22 &  0.30 
\end{tabular}
\caption{The octet and decuplet contributions to the octet baryon axial 
         couplings  both in dimensional
         regularization and for various values of the cutoff $\Lambda$ in MeV
         calculated with meson couplings at their $SU(6)$ values --- $D$=0.75,
         $F$=0.50, $C=-1.5$.}
\label{axtabdec}
\end{center}
\end{table}

\begin{table}[ht]
\begin{center}
\begin{tabular}{lrrrrr} 
      & Dim. & $\Lambda$=300 & $\Lambda$=400 & $\Lambda$=500 & $\Lambda$=600 \\
\hline
$g_A(\bar{p}n)$            &  0.92 &  0.20 &  0.28 &  0.37 &  0.45 \\
$g_A(\bar{p}\Lambda)$      & -0.95 & -0.18 & -0.27 & -0.36 & -0.45 \\
$g_A(\bar\Lambda\Sigma^-)$ &  0.62 &  0.12 &  0.18 &  0.23 &  0.29 \\
$g_A(\bar{n}\Sigma^-)$     &  0.19 &  0.04 &  0.05 &  0.07 &  0.09 \\
$g_A(\bar\Lambda\Xi^-)$    &  0.44 &  0.08 &  0.12 &  0.17 &  0.21 \\
$g_A(\bar\Sigma^0\Xi^-)$   &  1.14 &  0.21 &  0.31 &  0.42 &  0.53 
\end{tabular}
\caption{Octet contributions to the octet baryon axial 
         couplings  both in dimensional
         regularization and for various values of the cutoff $\Lambda$ in MeV
         calculated with meson couplings at their $SU(6)$ values --- $D$=0.75,
         $F$=0.50.}
\label{axtaboct}
\end{center}
\end{table}

In order to prove the chiral invariance of the cutoff procedure, we extract
the mass-independent divergent components of the integrals as $\Lambda$ goes 
to infinity, cf. Eqs.~(\ref{swaexp}) and (\ref{axexp}), which 
are compensated by the axial-vector couplings $D$ and $F$. By defining
\begin{eqnarray}
D^r &=& D + \frac{{\cal C}^2}{16 \pi^2 F_\pi^2} \left(  \frac{4}{3} \left[ 
     \frac{4}{3} D + 2F\right] \bar{J}^{{\rm div}}_\Lambda(\Delta)  
     - \frac{5}{18} [H + 9D] J^{{\rm div}}_\Lambda(\Delta) \right) \nonumber \\
F^r &=& F + \frac{{\cal C}^2}{16 \pi^2 F_\pi^2} \left(  \frac{40}{27}  
      D \bar{J}^{{\rm div}}_\Lambda(\Delta) 
     - \frac{5}{18} \left[\frac{5}{3}H + 9F\right]
 J^{{\rm div}}_\Lambda(\Delta) \right)
\end{eqnarray}
with 
\begin{eqnarray}
\bar{J}^{{\rm div}}_\Lambda(\Delta) &=& \frac{3}{4} \Lambda^2
                      - \frac{3\pi}{8}\Lambda \Delta
                      + \Delta^2\ln\Lambda + 2 \Delta^2 \nonumber \\
J^{{\rm div}}_\Lambda(\Delta) &=&  \Lambda^2  +4\Delta^2 \ln\Lambda 
                      - \pi \Delta \Lambda   + 2 \pi \Delta^2               
\end{eqnarray}
we are able to remove the divergences.
Similar to the case of the baryon masses we have also included
constant terms
$\Delta^2 $ and $2 \pi \Delta^2 $ which compensate effects
of the divergent  pieces, {\it e.g.}, 
$J^{{\rm div}}_{\Lambda = 380 {\rm MeV}}   \approx 0$, i.e. the numerical
value of the original integral is approximately maintained for the cutoff
range discussed here.
The mass-dependent divergences can also be renormalized by counterterms
of higher chiral orders, however the cumbersome calculation
involves numerous new counterterms\cite{nonlep} and is not presented here for brevity.

These asymptotic components of the integrals contain only short distance physics
and should therefore be removed from our final result. Thus we insert
the renormalized functions
\begin{equation}
\bar{J}^r_\Lambda(\Delta) = \bar{J}_\Lambda(\Delta) 
         - \bar{J}^{{\rm div}}_\Lambda(\Delta)  , \qquad
J^r_\Lambda(\Delta) = J_\Lambda(\Delta) - J^{{\rm div}}_\Lambda(\Delta)  
\end{equation}
into the expressions for the axial couplings
\begin{eqnarray} \label{axcut}
g_A [ij]  &=& - \frac{C^2}{24 \pi^2 F_\pi^2}\:
      \sum_{\phi=\pi,K,\eta} \: \bigg(  -  \frac{4}{3} 
         \rho_{ij}^{\, \phi} \bar{J}^r_\Lambda(m_\phi, \Delta)
              +\frac{5}{6}  H 
        \:\sigma_{ij}^{\, \phi} J^r_\Lambda(m_\phi, \Delta) \nonumber \\
  &&     + 
  \frac{3}{2} \gamma_{ij} (\epsilon_i^{\, \phi}  + \epsilon_j^{\, \phi} )
      J^r_\Lambda(m_\phi, \Delta)     \bigg)  
\end{eqnarray}
which yields the numerical results given in Tables \ref{axtabdec} and \ref{axtaboct}
($H=-2.0$). 
Different choices for $H$ do not change any of our conclusions, although the
actual values for the decuplet contributions are altered. But these changes are 
not significant since the contributions from the 
$H$-term in Eq.~(\ref{axcut}) are smaller than 
the remaining decuplet portions and the octet counterparts which do not depend on the
size of $H$.
In fact, for $\Lambda = 600$ MeV the contributions from the 
$H$-term are negligible. 
We conclude that the contributions from the loops are  smaller than in
the dim reg version and indicate that the convergence of the
chiral series is under control.

To check the quality of the HB$\chi$pt fits to experiment, the method
proposed in Subsection \ref{masscalc} indicates a theoretical uncertainty
of $(30\%)^3$ due to truncation of all HB$\chi$pt terms beyond $O(p^3)$.
A simple lowest order $SU(3)$ fit, which yields
$$D=0.80,\,\,\,F=0.46,\,\,\,\chi^2/d.o.f.=0.4,$$
is certainly a successful fit.
The results of including octet plus decuplet (octet only)
loop contributions are shown in Table 7 (Table 8) respectively.
Notice that LDR maintains the successful fit when loops are included,
since $\chi^2/d.o.f. \sim 1$, but dim reg produces a significantly poorer
fit.  For a complete calculation, there are extra counterterms that
should be added and in dim reg these will need to be large in order to
cancel the large loop effects indentified here.  In LDR, these subleading
counterterms will be small and no dramatic cancelations are required.

\begin{table}
\begin{tabular}{cccccc}
 &Dim.&$\Lambda$=300&$\Lambda$=400&$\Lambda=500$&$\Lambda$=600\\
\hline
$D$&0.78&0.85&0.77&0.73&0.70\\
$F$&-0.27&0.42&0.42&0.40&0.39\\
$\chi^2/d.o.f.$&2.2&0.8&1.0&1.4&1.8
\end{tabular}
\caption{Axial coupling fits with octet and decuplet loop contributions.
The weak axial decuplet 
parameter $H$ was chosen to have the value -2.0, as given in Ref.~\cite{ram}.}
\end{table} 

\begin{table}
\begin{tabular}{cccccc}

 &Dim.&$\Lambda$=300&$\Lambda$=400&$\Lambda=500$&$\Lambda$=600\\
\hline
$D$&0.55&0.70&0.68&0.65&0.63\\
$F$&0.31&0.41&0.39&0.37&0.36\\
$\chi^2/d.o.f.$&3.5&0.5&0.8&1.1&1.4
\end{tabular}
\caption{Axial coupling fits with octet loop contributions.}
\end{table}

\subsection{S-wave hyperon decays}
There exist seven nonleptonic hyperon decays: $\Sigma^+ \rightarrow n \pi^+$, 
$\Sigma^+ \rightarrow p \pi^0$, $\Sigma^- \rightarrow n \pi^-$, 
$\Lambda \rightarrow p \pi^-$, $\Lambda \rightarrow n \pi^0$, 
$\Xi^- \rightarrow \Lambda \pi^-$, and $\Xi^0 \rightarrow \Lambda \pi^0$.
Isospin symmetry of the strong interactions reduces the number of independent
amplitudes to four which we choose to be  $\Sigma^+ \rightarrow n \pi^+$, 
$\Sigma^+ \rightarrow p \pi^0$, $\Lambda \rightarrow n \pi^0$, 
 and $\Xi^0 \rightarrow \Lambda \pi^0$.

The general form of the decay amplitude is given by
\begin{equation}
{\cal A} =\sqrt{Z_a Z_b } 
    [ A^{\rm (tree)} + A^{\rm (loop)} ] ,
\end{equation}
where $Z_a$ and $Z_b$ denote the $Z$-factors of the incoming and 
outgoing baryon, respectively, and the amplitude has been decomposed
into tree and loop contributions. Let us focus on the decuplet loop
contributions.
They read
\begin{equation}
A^{\rm (loop)}  (B^a \rightarrow \tilde{B}^b \pi^i) = \frac{\sqrt{2}{\cal C}^2 h_c}{ 16 \pi^2 F_\pi^3}
           \sum_{\phi= \pi, K,\eta} \kappa(B^a_i)^\phi J(m_\phi,\Delta)
\end{equation}
with the coefficients
\begin{equation}
\begin{array}{lll}
\kappa(\Sigma^+_0)^\pi = -\frac{1}{9 \sqrt{2}}, &
\kappa(\Sigma^+_0)^K = -\frac{1}{18 \sqrt{2}}, &
\kappa(\Lambda^0_0)^\pi = -\frac{1}{2 \sqrt{3}},\\[0.3cm] 
\kappa(\Lambda^0_0)^K = -\frac{1}{4 \sqrt{3}}, &
\kappa(\Xi^0_0)^\pi = \frac{1}{4 \sqrt{3}}, &
\kappa(\Xi^0_0)^K = \frac{1}{4 \sqrt{3}}, 
\end{array}
\end{equation}
and all the remaining $\kappa(B^a_i)^\phi$ being zero.
In LDR the asymptotic mass-independent component of the 
integral $J_\Lambda(m,\Delta)$ can be absorbed by two counterterms at lowest order
\begin{equation}
{\cal L}_w = d_{w} \langle \bar{B} \{ h_+ , B\} \rangle + f_{w} \langle \bar{B}
[ h_+ , B] \rangle .
\end{equation}
A least-squares fit to the S-wave hyperon decays at tree level yields $d_{w}=0.16 \times
10^{-7}$ GeV and $f_{w}=-0.41 \times 10^{-7}$ GeV\cite{nonlep}.
The parameters $d_{w}$ and $f_{w}$ contribute to $A^{\rm (tree)}$ in the 
following way:
\begin{eqnarray}
A^{\rm (tree)} (\Sigma^+ \rightarrow p \pi^0) &=& \frac{1}{\sqrt{2}} \zeta(\Sigma^+_0) = 
             - \frac{1}{2 F_{\pi}} ( d_{w}-f_{w}), \nonumber \\
A^{\rm (tree)} (\Lambda \rightarrow n \pi^0) &=& \frac{1}{\sqrt{2}} \zeta(\Lambda^0_0) =
             \frac{1}{2 \sqrt{6} F_{\pi}} ( d_{w}+3f_{w}), \nonumber \\
A^{\rm (tree)} (\Xi^0 \rightarrow \Lambda \pi^0) &=& \frac{1}{\sqrt{2}} \zeta(\Xi^0_0) =
                 \frac{1}{2 \sqrt{6} F_{\pi}} ( d_{w}-3f_{w}), \nonumber \\
A^{\rm (tree)} (\Sigma^+ \rightarrow n \pi^+) &=& \frac{1}{\sqrt{2}} \zeta(\Sigma^+_+) =0 .
\end{eqnarray}
The divergent pieces of the prefactor $\sqrt{Z_a Z_b }$ must be
taken into account as well. We note that the asymptotic pieces for all decays
are given by
\begin{equation}
\left. \sqrt{Z_a Z_b } \,\right| _{\rm asy} = - \frac{5{\cal C}^2 }{ 32 \pi^2 F_\pi^2}
              \left[ \Lambda^2 - \pi \Delta \Lambda + 4 \Delta^2 \ln \Lambda   \right] 
\end{equation}
so that one can immediately absorb these components into a redefinition
of $d_{w}$ and $f_{w}$.

By defining the renormalized parameters
\begin{eqnarray}
d^{r}_{w} &=& d_{w} + \frac{{\cal C}^2 }{ 32 \pi^2 F_\pi^2} (h_c -5 d_{w})
\left[ \Lambda^2 - \pi \Delta \Lambda + 4 \Delta^2 \ln \Lambda + 2 \pi \Delta^2 \right],
   \nonumber \\
f^{r}_{w} &=& f_{w} + \frac{5{\cal C}^2 }{ 96 \pi^2 F_\pi^2} (h_c -3 f_{w})
\left[ \Lambda^2 - \pi \Delta \Lambda + 4 \Delta^2 \ln \Lambda + 2 \pi \Delta^2 \right],    
\end{eqnarray}
we are able to remove the short distance portion of both $A^{\rm (loop)} $
and the $Z$-factors. The mass-dependent divergences can be removed in a similar way,
but the calculation is rather involved and we refrain from presenting it here. 
(For the complete renormalization at one-loop order in dim reg see Refs.~\cite{nonlep, renorm}.)

Having assured ourselves about the conservation of chiral invariance for the
S-wave hyperon decays also, we calculate numerically the decuplet
contributions to the decays by employing the formula
\begin{equation}  \label{nonlepform}
\delta {\cal A} (B^a \!\rightarrow \tilde{B}^b \pi^i) =  \frac{\sqrt{2}{\cal C}^2 }{
16 \pi^2 F_\pi^3}    \sum_{\phi= \pi, K,\eta}
    \left[ h_c \kappa(B^a_i)^\phi - \frac{1}{2} \zeta(B^a_i) 
     (\epsilon_B^\phi + \epsilon_{\tilde{B}}^\phi ) \right]
     J^r_\Lambda(m_\phi, \Delta) .
\end{equation}
The numerical results are given in Tables \ref{nleptabdec} and
\ref{nleptaboct} where we used
$d_{w}=0.16 \times 10^{-7}$ GeV and $f_{w}=-0.41 \times 10^{-7}$ GeV.
The convergence of the chiral series is again improved by using LDR. 
The results are afflicted with  some uncertainty due to the appearance
of a new parameter 
$h_c$  which was obtained with large error bars
from a fit to data using dim reg, $h_c = (-0.10 \pm 0.19) \times 10^{-7}$\cite{jen}.
Changes in $h_c$  affect the numerical results only slightly, 
in particular for a cutoff of $\Lambda = 600$ MeV the
contributions from the $h_c$ term in Eq.~(\ref{nonlepform}) are in general negligible.
Our conclusions are therefore independent of the particular choice for $h_c$.

\begin{table}[ht]
\begin{center}
\begin{tabular}{lrrrrr}  
      & Dim. & $\Lambda$=300 & $\Lambda$=400 & $\Lambda$=500 & $\Lambda$=600 \\
\hline
$\delta {\cal A} (\Lambda_0^0)$ & 1.12 &  0.42 & -0.09 & -0.53  & -0.91  \\
$\delta {\cal A} (\Xi_0^0)$     & -3.75 & -0.70 & -0.15 & 0.30  & 0.70  \\
$\delta {\cal A} (\Sigma_0^+)$  & 3.70 & 0.75 & 0.25 & -0.16  & -0.52 
\end{tabular}
\caption{Shown are the octet and decuplet contributions to the S-wave hyperon decays
         in units of $10^{-7}$ both in dimensional
         regularization and for various values of the cutoff $\Lambda$ in MeV.}
\label{nleptabdec}
\end{center}
\end{table}

\begin{table}[ht]
\begin{center}
\begin{tabular}{lrrrrr}  
      & Dim. & $\Lambda$=300 & $\Lambda$=400 & $\Lambda$=500 & $\Lambda$=600 \\
\hline
$\delta {\cal A} (\Lambda_0^0)$ & -1.87  & -0.32 & -0.50 & -0.68 & -0.87 \\
$\delta {\cal A} (\Xi_0^0)$     &  1.00  &  0.18 &  0.28 &  0.38 &  0.47 \\
$\delta {\cal A} (\Sigma_0^+)$  & -0.72  & -0.12 & -0.19 & -0.26 & -0.33
\end{tabular}
\caption{Octet contributions to the S-wave hyperon decays
         in units of $10^{-7}$ both in dimensional
         regularization and for various values of the cutoff $\Lambda$ in MeV.}
\label{nleptaboct}
\end{center}
\end{table}

The omission of $O(p^4)$ HB$\chi$pt contributions leads to a fractional
theoretical uncertainty of $(30\%)^3$ = 0.03.  Combining this with the
experimental uncertainties leads to the following lowest order $SU(3)$ fit
(here $d_w,f_w$ are in units of $10^{-7}$ GeV):
$$d_w=0.18,\,\,\, f_w=-0.41,\,\,\,\chi^2/d.o.f.=1.2.$$
The effect of including chiral loop corrections with octet plus decuplet
(octet only) states is shown in Table 11 (Table 12).
In the case of LDR, the $\chi^2/d.o.f.$ values are close to being
acceptable, and the fit becomes perfectly fine when the $O(p^3)$ counterterms,
omitted here, are added in.  In the dim reg case, addition of the decuplet
loops severely worsens the fit and the missing counterterms must be assigned
huge values if the fit is to be repaired.  This is another signal of the
breakdown of convergence for dim reg.

\begin{table}
\begin{tabular}{cccccc}
 &Dim.&$\Lambda$=300&$\Lambda$=400&$\Lambda=500$&$\Lambda$=600\\
\hline 
$d_w$&-1.65&0.26&0.22&0.20&0.18\\
$f_w$&0.55&-0.52&-0.42&-0.36&-0.32\\
$\chi^2/d.o.f.$&502&1.9&2.2&2.4&2.6
\end{tabular}
\caption{S-wave hyperon decay fits with octet plus decuplet 
loop contributions and
calculated with lowest order meson couplings having their $SU(6)$ 
values --- $D$=0.75, $F$=0.50, $C=-1.5$.  Here the decuplet weak decay
parameter $h_c=-0.1\times 10^{-7}$ was used, as recommended in Ref.~\cite{jen},
and $d_w$ and $f_w$ are in units of $10^{-7}$. }
\end{table} 
 
\begin{table}
\begin{tabular}{cccccc}

 &Dim.&$\Lambda$=300&$\Lambda$=400&$\Lambda=500$&$\Lambda$=600\\
\hline
$d_w$&0.20&0.19&0.20&0.20&0.20\\
$f_w$&-0.28&-0.38&-0.36&-0.35&-0.34\\
$\chi^2/d.o.f.$&11&2.0&2.6&3.4&4.3
\end{tabular}
\caption{S-wave hyperon decay fits with octet loop contributions and
calculated with lowest order meson couplings having their $SU(6)$ 
values --- $D$=0.75, $F$=0.50.
$d_w$ and $f_w$ are in units of $10^{-7}$. }
\end{table} 

\subsection{Magnetic moments}
Finally, we turn to the calculation of the magnetic moments.
The decuplet loop contributions to the octet baryon magnetic moments
have been found to improve the convergence of the chiral series
in dim reg \cite{prm}. They
have the structure
\begin{equation}
\delta \mu_i = \frac{M_0 {\cal C}^2}{ 144 \pi^2 F_\pi^2} 
         \sum_{\phi= \pi, K} \lambda_i^\phi K(m_\phi, \Delta) 
\end{equation}
with the coefficients
\begin{equation}
\begin{array}{llll}
\lambda_{p}^\pi = 4 ,   & \lambda_{p}^K = -1 , 
& \lambda_{n}^\pi = -4 ,   & \lambda_{n}^K = -2 ,\\[0.2cm]
\lambda_{\Sigma^+}^\pi = -1 ,   & \lambda_{\Sigma^+}^K = 4 ,
& \lambda_{\Sigma^-}^\pi = 1 ,   & \lambda_{\Sigma^-}^K = 2 ,\\[0.2cm]
\lambda_{\Sigma^0}^\pi = 0 ,   & \lambda_{\Sigma^0}^K = 3,
& \lambda_{\Xi^-}^\pi = 2 ,   & \lambda_{\Xi^-}^K = 1 ,\\[0.2cm]
\lambda_{\Xi^0}^\pi = -2 ,   & \lambda_{\Xi^0}^K = -4 ,
& \lambda_{\Lambda^-}^\pi = 0 ,   & \lambda_{\Lambda^-}^K = -3 ,\\[0.2cm]
\lambda_{\Lambda \Sigma^0}^\pi = 2 \sqrt{3} ,  & \lambda_{\Lambda \Sigma^0}^K = \sqrt{3}  .\qquad  
&&
\end{array}
\end{equation}
Omitting the terms in $K(m, \Delta)$ proportional to $\tilde{L}$ we obtain the
numerical results in Tables \ref{magtabdec} and \ref{magtaboct},
where we used $M_0 =767$ MeV\cite{mbm}.

\begin{table}[ht]
\begin{center}
\begin{tabular}{lrrrrr}  
      & Dim. & $\Lambda$=300 & $\Lambda$=400 & $\Lambda$=500 & $\Lambda$=600 \\
\hline
$\delta \mu_p$               & -2.43 & -0.71 & -0.82 & -0.91 & -1.00 \\
$\delta \mu_n$               &  1.36 &  0.45 &  0.46 &  0.46 &  0.47 \\
$\delta \mu_\Lambda$         &  1.88 &  0.31 &  0.36 &  0.42 &  0.47 \\
$\delta \mu_{\Sigma^+}$      & -3.83 & -0.79 & -0.96 & -1.11 & -1.25 \\
$\delta \mu_{\Sigma^0}$      & -1.88 & -0.31 & -0.36 & -0.42 & -0.47 \\
$\delta \mu_{\Sigma^-}$      &  0.07 &  0.18 &  0.24 &  0.28 &  0.31 \\
$\delta \mu_{\Sigma\Lambda}$ & -1.79 & -0.44 & -0.47 & -0.50 & -0.54 \\
$\delta \mu_{\Xi^-}$         &  1.21 &  0.26 &  0.37 &  0.46 &  0.54 \\
$\delta \mu_{\Xi^0}$         &  3.62 &  0.61 &  0.72 &  0.83 &  0.93
\end{tabular}
\caption{Shown are the octet and decuplet contributions to the octet baryon 
         magnetic moments both in dimensional
         regularization and for various values of the cutoff $\Lambda$ in MeV.}
\label{magtabdec}
\end{center}
\end{table}

\begin{table}[ht]
\begin{center}
\begin{tabular}{lrrrrr}  
      & Dim. & $\Lambda$=300 & $\Lambda$=400 & $\Lambda$=500 & $\Lambda$=600 \\
\hline
$\delta \mu_p$               & -2.29 & -0.67 & -0.81 & -0.93 & -1.03 \\
$\delta \mu_n$               &  0.65 &  0.35 &  0.41 &  0.44 &  0.47 \\
$\delta \mu_\Lambda$         &  1.31 &  0.25 &  0.32 &  0.39 &  0.44 \\
$\delta \mu_{\Sigma^+}$      & -3.16 & -0.73 & -0.91 & -1.07 & -1.21 \\
$\delta \mu_{\Sigma^0}$      & -1.31 & -0.25 & -0.32 & -0.39 & -0.44 \\
$\delta \mu_{\Sigma^-}$      &  0.54 &  0.23 &  0.27 &  0.30 &  0.32 \\
$\delta \mu_{\Sigma\Lambda}$ & -1.18 & -0.35 & -0.43 & -0.49 & -0.54 \\
$\delta \mu_{\Xi^-}$         &  1.56 &  0.31 &  0.39 &  0.47 &  0.54 \\
$\delta \mu_{\Xi^0}$         &  2.70 &  0.51 &  0.65 &  0.78 &  0.90
\end{tabular}
\caption{Octet contributions to the octet baryon 
         magnetic moments both in dimensional
         regularization and for various values of the cutoff $\Lambda$ in MeV.}
\label{magtaboct}
\end{center}
\end{table}

Before comparing with the cutoff version, we prove the chiral invariance of the
cutoff procedure.
There exist two counterterms at lowest order\cite{mml}
\begin{equation}
{\cal L} = - \frac{i}{4 M_0} b_6^D \langle \bar{B} [S^\mu, S^\nu] \{ f_{\mu \nu}^+, B\} \rangle
     - \frac{i}{4 M_0} b_6^F \langle \bar{B} [S^\mu, S^\nu] [ f_{\mu \nu}^+, B] \rangle
\end{equation}
where 
\begin{equation}
f_{\mu \nu}^+ = u^\dagger f_{\mu \nu} u + u f_{\mu \nu} u^\dagger 
            =  -2 e Q [ \partial_\mu {\cal A}_\nu - \partial_\nu {\cal A}_\mu ]
             + {\cal O}(\phi^2)
\end{equation}
is the field strength tensor  for the photon field ${\cal A}_\mu$
 and $Q= \frac{1}{3}{\rm diag}(2,-1,-1)$ the quark charge matrix.           
By introducing the renormalized parameters
\begin{eqnarray}
b_6^{D \, ; \, r} &=& b_6^D - \frac{M_0 {\cal C}^2}{48 \pi^2 F_\pi^2} ( \pi \Lambda
      - 6 \Delta \ln \Lambda - 3 \pi \Delta), \nonumber \\
b_6^{F \, ; \, r} &=& b_6^F,
\end{eqnarray}
we are able to remove the asymptotic pieces of $K_\Lambda(m, \Delta)$.
Our final result reads
\begin{equation}
\delta \mu_i = \frac{M_0 {\cal C}^2}{ 144 \pi^2 F_\pi^2} 
         \sum_{\phi= \pi, K} \lambda_i^\phi K_\Lambda^r(m_\phi, \Delta) 
\end{equation}
with 
\begin{equation}
K_\Lambda^r(m, \Delta) = K_\Lambda(m, \Delta) - \frac{\pi}{3} \Lambda + 2 \Delta \ln \Lambda 
+ \pi \Delta
\end{equation}
where the constant piece $\pi \Delta$ compensates approximately the numerical effects
of the $\Lambda$-dependent terms in the region 300 MeV $<\Lambda<$ 600 MeV,
and the final results can be found in Tables  \ref{magtabdec} and \ref{magtaboct}.
Again, the loop contributions yield smaller results in LDR than in dim reg
with the only exception being $\delta \mu_{\Sigma^-}$
where there happen to be large cancelations between the octet and decuplet
contributions in dim reg.

We can study the goodness of fit as done above for the other
observables.  In this case the HB$\chi$pt expressions begin at $O(p^2)$
so the relative theoretical error is $(30\%)^2$.  A lowest order $SU(3)$
fit yields $$b_6^D=2.46,\,\,\,b_d^F=1.76,\,\,\,\chi^2/d.o.f.=2.3$$
while the effects of loop corrections with octet plus decuplet (octet only)
states are shown in Table 15 (Table 16).

\begin{table}
\begin{tabular}{cccccc}

 &Dim.&$\Lambda$=300&$\Lambda$=400&$\Lambda=500$&$\Lambda$=600\\
\hline
$b_6^d$&6.15&3.27&3.35&3.43&3.51\\
$b_6^F$&3.67&2.23&2.35&2.44&2.53\\
$\chi^2/d.o.f.$&39&5.1&3.6&2.4&1.5
\end{tabular}
\caption{Magnetic moment fits with octet and decuplet loop contributions and
calculated with lowest order meson couplings having their $SU(6)$ 
values --- $D$=0.75, $F$=0.50, $C=-1.5$.}
\end{table} 

\begin{table}
\begin{tabular}{cccccc}

 &Dim.&$\Lambda$=300&$\Lambda$=400&$\Lambda=500$&$\Lambda$=600\\
\hline
$b_6^d$&4.92&3.11&3.26&3.38&3.49\\
$b_6^F$&3.67&2.23&2.35&2.44&2.53\\
$\chi^2/d.o.f.$&23&5.0&4.0&2.9&2.0
\end{tabular}
\caption{Magnetic moment fits with octet loop contributions and
calculated with lowest order meson couplings having their $SU(6)$ 
values --- $D$=0.75, $F$=0.50.}
\end{table} 

All of the $\chi^2/d.o.f.$ values are noticeably larger than unity,
suggesting that higher orders of HB$\chi$pt are required.  This is
not too surprising since leading order for the magnetic moments
is $O(p^2)$ rather than $O(p)$.  It should be noted, however, that
LDR maintains the moderate $\chi^2/d.o.f.$ that was obtained from the
tree-level fit, whereas the huge loop effects in dim reg destroy the
fit completely.

\section{Conclusions}
In this work we have
investigated the use of long distance regularization, LDR, in baryon $\chi$pt
with decuplet fields. For practical purposes we employed a dipole regulator as a cutoff,
however the specific shape of the cutoff is irrelevant as long as it preserves 
chiral symmetry.

We extended our previous LDR methods to include the case of the decuplet by
examining the loop contributions to the octet baryon masses, axial couplings, 
S-wave nonleptonic hyperon decays and magnetic moments.
In each of these four cases we were able to show that 
all possible power divergences (modulo logs)
in the cutoff could be removed by redefining the 
coupling constants of the Lagrangian.
This verifies the chiral consistence of the cutoff procedure --- a chiral
expansion {\it can} be carried out to the order we are working.
By choosing a finite cutoff in the range 300 MeV $< \Lambda <$ 600 MeV,
which corresponds to distances of about the baryon radius, we are able to
separate the low momentum {\it long distance} physics from the high
momentum {\it short distance} portion of the
integrals.  The key point here is that it is this long distance
component, whose form is determined simply by the basic chiral
symmetry of the underlying QCD Lagrangian, which can be trusted.  Once
one is dealing with interactions at a scale of order the size of the
baryon or smaller, structure issues surely significantly modify the 
form and results of the loop integrals.  Without a detailed model, 
we do not know the form that such changes take, so our approach is to use 
a cutoff scheme, wherein such short distance effects are suppressed.  
While this approach is a bit brute force in nature, the results are
quite consistent with detailed model calculations such as found in a
Bethe-Salpeter approach\cite{kre} or in the cloudy bag
model\cite{clb}.  
In any case it is a general property that within this
procedure chiral invariance is maintained but smaller loop effects are
found than in the conventional dim reg method,
leading in general therefore to less damage to the traditional $SU(3)$ fits to 
these processes.  We studied the goodness of fit in each case and
found that dim reg at one
chiral loop often gave a substantially degraded fit, which was
improved by the LDR procedure.

This work also shed some light on the interplay between the chiral and
1/$N_c$ limits, since octet and decuplet fields become degenerate as 
$N_c \rightarrow \infty$.  Indeed the success of $1/N_c$ arguments in
other contexts {\it demands} the addition of decuplet effects to these
and other baryonic applications.  In general, we found that such
inclusion could be handled straightforwardly and that the resulting
changes to our previous octet-only loop results were relatively minor.
   
Moreover, it has been emphasized that LDR ideas are
useful when extrapolating lattice data down to physical pion masses\cite{lei}
and the present investigation may help to compare with such
chiral extrapolations of lattice calculations which are 
usually done with a pion mass in the vicinity of 600 MeV or so.  
Our study is a first step towards comparing and combining
these different techniques and will hopefully improve the communication
between the involved communities of physicists.

\begin{center}
{\bf\large Acknowledgements}
\end{center}

B.B. is grateful to the High Energy Theory Group of the
University of Massachusetts Amherst for hospitality during the later 
stages of this work.
The work of B.R.H. was partially supported by the National Science
Foundation under award PHY-9801875, that of R.L. and P.-P.A.O. by
the Natural Sciences and Engineering Research Council of Canada
and that of B.B. by the Deutsche Forschungsgemeinschaft.


\end{document}